\documentclass{article}
\usepackage{dcase2019,amsmath,graphicx,url,times,booktabs, tabularx}
\usepackage{multirow}
\usepackage{algorithm}
\usepackage{algpseudocode}
\usepackage{caption}
\usepackage{subcaption}

\title{A TWO-STEP SYSTEM FOR SOUND EVENT LOCALIZATION AND DETECTION}

\name{Thi Ngoc Tho Nguyen$^{1}\sthanks{This material is based on research work supported by the IAF-ICP: Singtel Cognitive and Artificial Intelligence Lab for Enterprises@NTU under the Research Theme on Edge Intelligence.}$,
       Douglas L. Jones$^{2}$, 
       Rishabh Ranjan $^{3}$,
       Sathish Jayabalan$^{3}$,
       Woon Seng Gan$^{1}$,
       }
\address{$^1$ Nanyang Technological University, Electrical and Electronic Engineering Dept., Singapore,\\
 			 \{nguyenth003, ewsgan\}@ntu.edu.sg\\          
         $^2$ University of Illinois at Urbana-Champaign, Dept. of Electrical and Computer Engineering,
         \\ Illinois, USA, \{dl-jones\}@illinois.edu\\
         $^3$ Nanyang Technological University, SCALE, Singapore,\\
 			 \{rishabh001, sathishj\}@ntu.edu.sg\\  
}

\begin{document}

\ninept
\maketitle

\begin{sloppy}

\begin{abstract}
Sound event detection and sound event localization requires different features from audio input signals. While sound event detection mainly relies on time-frequency patterns to distinguish different event classes, sound event localization uses magnitude or phase differences between microphones to estimate source directions. Therefore, we propose a two-step system to do sound event localization and detection. In the first step, we detect the sound events and estimate the directions-of-arrival separately. In the second step, we combine the results of the event detector and direction-of-arrival estimator together. The obtained results show a significant improvement over the baseline solution for sound event localization and detection in DCASE 2019 task 3 challenge. Using the evaluation dataset, the proposed system achieved an F$1$ score of $93.4$\% for sound event detection and an error of $5.4^{\circ}$ for direction-of-arrival estimation, while the winning solution achieved an F$1$ score of $94.7\%$ and an angle error of $3.7^{\circ}$ respectively.  
\end{abstract}


\section{Introduction}
\label{sec:intro}

Sound event localization and detection (SELD) has a wide application in acoustic monitoring, and context-aware devices~\cite{takeda2016sound, Adavanne2018_JSTSP}. SELD can provide the information of sound classes and the corresponding directions-of-arrival (DOAs) of multiple sound sources. As a result, SELD is the core component of acoustic monitoring systems such as environmental noise monitoring, and surveillance system. For example, SELD can direct a surveillance camera to point toward the direction of some sounds of interest. In addition, SELD can also assist context-aware devices such as hearing aids, smart phones, autonomous cars, and robots to be aware of the surrounding environments. 

DCASE 2019 task 3, sound event localization and detection, challenges participants to detect sound events and their corresponding directions-of-arrival ~\cite{Adavanne2019_DCASE}. There are a total of 11 sound classes taken from DCASE 2016 task $2$ dataset. The clean data are convolved with real-file recorded room impulse responses from $5$ different indoor locations. $50$\% of the synthesized clips has $2$ temporal overlapping sound events. The room impulse responses are recorded using Eigenmike microphone array at every $10^{\circ}$ azimuth angle between $-180^{\circ}$ and $180^{\circ}$, and at every $10^{\circ}$ elevation between $-40^{\circ}$ and $40^{\circ}$. The data are given in two formats: first-order ambisonics and tetrahedral microphone array.   

In general, SELD consists of two components, which are sound source localization (SSL) and sound event detection (SED). A microphone array is required to do sound source localization. In the context of the DCASE 2019 task 3 challenge, SSL refers to DOA estimation. The main challenges of SED tasks are multiple sources, overlapping events, varying background noises, and lacking of labeled data. In the past decade, deep learning is the most common solution for SED tasks~\cite{virtanen2018computational}. The state-of-the-art SED models are often learnt using convolutional neural networks (CNN)~\cite{Salamon2017Cnn, piczak2015environmental}, recurrent neural networks (RNN)~\cite{parascandolo2016recurrent}, and some combinations of these two networks such as convolutional recurrent neural networks (CRNN)~\cite{Adavanne2018_JSTSP, cakir2017convolutional}. The main challenges of DOA estimation are reverberation, multiple sources, and varying background noises. Traditionally, DOA tasks for small microphone arrays are solved by using signal processing algorithms such as minimum variance distortionless response (MVDR) beamformer~\cite{Capon1969mvdr}, multiple signal classification (MUSIC)~\cite{schmidt1986multiple}, and steered response power (SRP)~\cite{salvati2014incoherent}. Recently, several researches have applied deep learning to DOA estimations~\cite{xiao2015learning, adavanne2018doaRcnn, he2018deep}. Compared to the signal processing methods, the deep learning methods require more data for training, and the models need to be retrained when another microphone configuration is used. 

There are two main approaches to solve for SELD. One is the end-to-end approach~\cite{Adavanne2018_JSTSP} where one system learns to detect the sound events and estimate their DOAs simultaneously. The end-to-end approach is attractive since it associate the sound events and DOA estimates explicitly. However, SED and DOA estimation require different types of information and thus a joint estimation can hurt the performance of the whole SELD system. The baseline solution in DCASE task 3 is a joint CRNN model for SED and DOA estimation. The baseline model inputs both the magnitude and phase spectrograms of all microphone channels, and do multi-label classification for SED and regression for DOAs. The resulting DOA error on the development dataset is relatively large at $28.5^{\circ}$, and the SED F$1$ score is $79.9$\% for the first-order ambisonic format. To mitigate this problem, Cao \emph{et al.}~\cite{cao2019polyphonic} proposed a two-stage strategy to train a SELD network. First, the SED branch is trained using all the available data. After that, the weight of the CNN portion of the SED branch is transferred to the DOA branch. The DOA branch is trained using only the data that have active sources. The two-stage SELD network inputs log-mel spectrogram and GCC-PHAT features. This training scheme reduced the DOA error on the development dataset significantly to $9.84^{\circ}$, and improved the SED F$1$ score to $89.8$\%. 

The other approach is to solve for SED and DOA separately, and match the sound events with the DOA estimates later. In order to maximize the performance of the two subtasks, we propose a two-step SELD system, where we detect the sound events and estimate their DOAs separately and joint them later. We use a parametric-based algorithm for DOA estimation and deep learning model for SED. The advantage of this approach is that it does not require training data for DOA estimation. As a result, this approach can be applied to many array configurations where a required SELD dataset is not available. The main drawback of the two-step system is that mismatches of the sound events and their DOAs occur when there are overlapping events. The obtained results show a significant improvement over the joint estimation proposed in the baseline system. We achieved the best DOA error among all the proposed parametric approaches in DCASE 2019 task 3 challenge. Using the development dataset, our ensemble model achieved a DOA error of $5.12^{\circ}$ and a SED F$1$ score of $89.3$\%. Using the evaluation dataset, our ensemble model achieved a DOA error of $5.4^{\circ}$ and a SED F$1$ score of $93.4$\%. We ranked $6^{th}$ in the team ranking category. We organize the paper as follows. Section II shows our proposed two-step system for SELD task. Section III presents experimental results and discussions. Finally, we conclude the paper in Section IV. 

\section{A two-step SELD system}
\label{sec:format}

We use the first-order ambisonic (FOA) format for the SELD task. The development set consists of $400$ one-minute audio clips divided into $4$ folds. The evaluation set consists of $100$ one-minute audio clips. Microphone input signals are transformed into the short-time Fourier transform domain with the following parameters: sampling rate of $48$ kHz, window length of $2048$ samples, hop length of  $960$ samples ($0.02$ second), and $2048$ FFT points. This results in $3000$ time frames for each one-minute audio clip. We predict the classes of sound events and their corresponding DOAs for each of the $3000$ time frame. The block diagram of our two-step SELD system is shown in Fig~\ref{fig1:seld_system}. We use convolutional recurrent neural network (CRNN) for SED, and a single-source histogram algorithm~\cite{tho2014robust} for DOA estimation. The results of SED and DOA estimation are fused together for each time frame using rule-based logics. 

\begin{figure}[tb]
\centering
\includegraphics[width=0.3\textwidth]{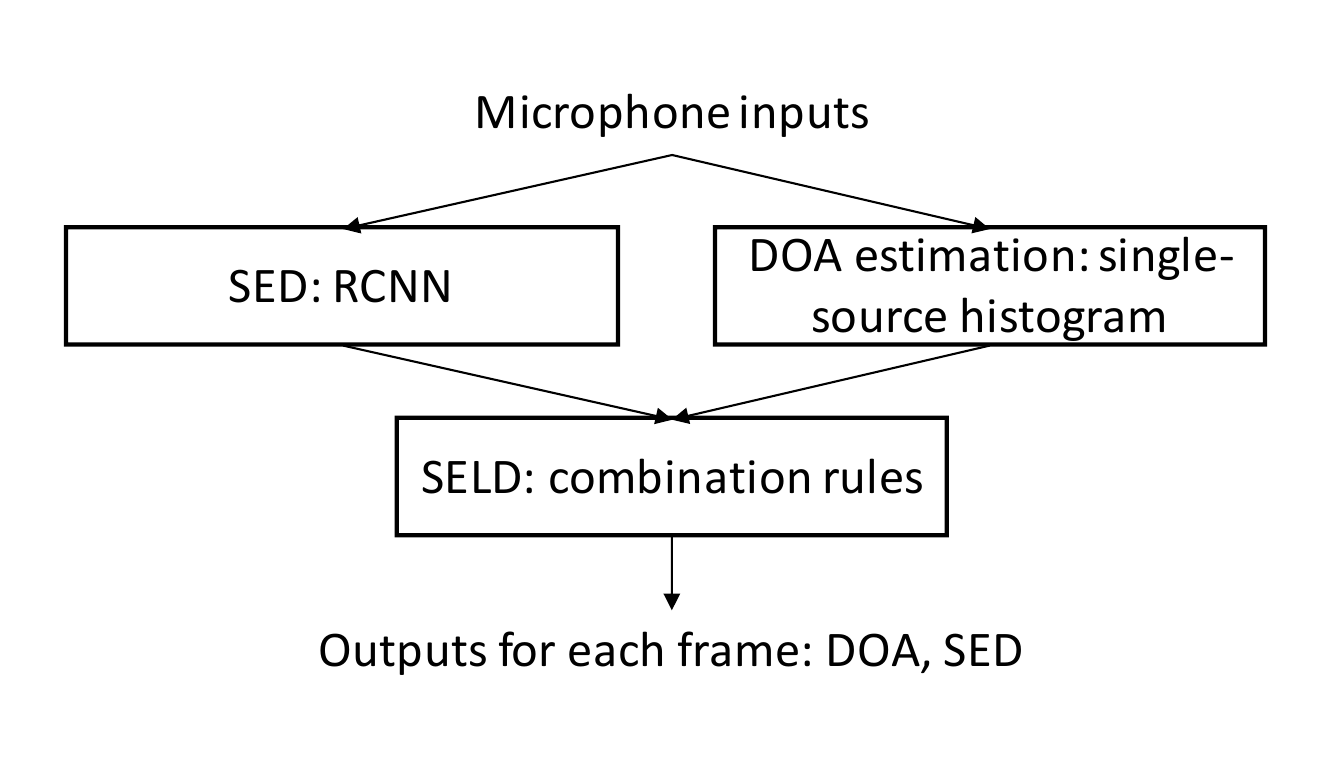}
\caption{A two-step SELD system}
\label{fig1:seld_system}
\end{figure} 

\subsection{Sound event detection}
\label{sec:sed}

We extract $128$ log mel-band energies of all the $4$ microphones as input features for the SED block. The audio signals are divided into $128$ frame segments. The size of input features into CRNN model is $128$ frame x $128$ mel x $4$ channels. We replace the $3$ convolutional layers in the baseline model~\cite{Adavanne2019_DCASE} with the first three stages of Resnet-$50$ network~\cite{he2016deep}. Since the size of the SELD dataset is much smaller compared to the ImageNet~\cite{deng2009imagenet} that are used to train the original Resnet-$50$, we reduce the number of filters in each stage. Table.~\ref{table1:SED_network} shows the SED network architecture. We will refer to this network in the subsequent sections as the CRNN-Resnet. The log-mel input features are normalized along the mel-band by mean and standard deviation. We use the same normalization factors for all $4$ channels. Throughout the network, we do max pooling only on the mel-band dimension, so that we can have one output prediction at each time frame. 

In the first stage, the input features are fed into a convolutional layer with $32$ filters of size $5$x$5$, batch normalization, ReLU activation, and max pooling of size $1$ x $4$. The second and fourth stages are the modified stage $2$ and $3$ of the Resnet-$50$ network respectively. We reduce the number of filters to [$32$,$32$,$128$] for stage $2$, and [$64$,$64$,$256$] for stage $4$. We insert two average pooling layers of stride ($1$, $4$) to reduce the number of parameters and avoid overfitting. The subsequent recurrent layers and fully connected layers are similar to the baseline~\cite{Adavanne2019_DCASE}. The output layer uses sigmoid activation to do multi-label classification. 

We train the network using Adam optimizer with learning rate of $0.0001$ for $100$ epochs. Cross validation is used to select the best parameters to train the final model. The competition uses individual evaluation metrics for SED and DOA estimation. For SED task, evaluation metrics are error rate and F1 score calculated in one-second segment~\cite{Mesaros2016_MDPI}. To make our predictions more robust against the random segmentation of the audio clip, for the validation and testing data, we shift each of the audio clip $0$, $32$, and $64$ samples before dividing them into $128$-frame segments, and input them into the SED network. After that we combine the $3$ predictions of the SED network at each frame using geometric mean. On the $4$-fold cross validation, this shifting scheme reduces our error rate about $0.06$ and improves the F$1$ score of $1$\%. In the final prediction, a sound event is consider active if the prediction probability is greater than $0.5$. 

\begin{table}\small
\centering
\caption {SED network architecture}  \vspace*{5pt}
\label{table1:SED_network}
\scalebox{0.80}{
\begin{tabular}{|c|c|c|}
\hline 
Stage & Output & Layers \\ 
\hline 
$1$ & $128$x$64$x$32$ & 32 5x5 conv, BN, ReLU, maxpool, stride = ($1$,$2$) \\ 
\hline 
\multirow{3}{*}{\centering $2$}   		      &
\multirow{3}{*}{\centering $128$x$64$x$128$}  & conv block, filters =[$32$,$32$,$128$], stride = ($1$,$1$)\\ 
											  \cline{3-3}
											  & & identity block, filters =[$32$,$32$,$128$] \\ 
											  \cline{3-3}
											  & & identity block, filters =[$32$,$32$,$128$] \\ 
											  \cline{3-3}
\hline
$3$ & $128$x$16$x$128$ & Average pooling, stride = ($1$,$4$) \\ 
\hline 
\multirow{3}{*}{\centering $4$}   		      &
\multirow{3}{*}{\centering $128$x$8$x$128$}  & conv block, filters =[$64$,$64$,$256$], stride = ($1$,$2$) \\ 
											  \cline{3-3}
											  & & identity block, filters =[$64$,$64$,$256$] \\ 
											  \cline{3-3}
											  & & identity block, filters =[$64$,$64$,$256$] \\ 
											  \cline{3-3}
\hline 
$5$ & $128$x$2$x$256$ & Average pooling, stride = ($1$,$4$) \\ 
\hline 
$6$ & $128$x$512$ & Reshape \\ 
\hline 
$7$ & $128$x$128$  & Bidirectional, 128 GRU units, tanh  \\ 
\hline 
$8$ & $128$x$128$  & Bidirectional, 128 GRU units, tanh  \\  
\hline 
$9$ &  $128$x$128$ & $128$ fully connected \\ 
\hline 
$10$ &  $128$x$11$ & $11$ fully connected, sigmoid activation \\ 
\hline 
\multicolumn{2}{|c|}{Number of parameters} & $1008235$ \\
\hline 
\end{tabular} 
}
\end{table}

We experiment with several variations of the CRNN-Resnet architecture. The first variation uses 128 log-mel energy of background-normalized magnitude spectrograms that have been used in our previous publication~\cite{Nguyen2018b}. The magnitude spectrogram is normalized by a noise floor to mitigate the effect of different background noise levels. The background-normalized spectrogram does not perform as well as the non-normalized version but it often helps the ensemble model. The second variation uses LSTM instead of GRU in the recurrent layers. The third variance uses additional inputs which is the largest eigenvector of the covariance matrix of each TF bin; the fourth variance has additional output which indicates if a time frame has an active signal. We combine the CRNN-Resnet model and its four variations into an ensemble to improve the overall SELD results.

\subsection{DOA estimation}
\label{sec:doa_est}

We use a single-source histogram algorithm proposed in ~\cite{tho2014robust} to estimate DOAs. The single-source histogram finds all the time-frequency (TF) bins that contains energy from mostly one source. A TF bin is considered to be a single-source TF bin when it passes all three tests: magnitude, onset, and coherence test. Magnitude test finds TF bins that are above a noise floor to mitigate the effect of background noise. Onset test finds TF bins that belong to direct-path signals to reduce the effect of reverberation in the DOA estimation. Coherence test finds TF bins of which the covariance matrices are approximately rank-$1$. After all the single-source TF bins are found, the DOA at each bin is computed using the theoretical steering vector of the microphone array~\cite{tho2014robust}. These DOAs are discretized using the required resolution of azimuth and elevation angles. After that, these DOAs are populated into a histogram. The histogram is smoothed to reduce the estimation errors. The final DOA estimates are the peaks of this histogram. 

We compute one histogram per time frame. Since the SELD dataset have only maximum two overlapping sources and moderate level of reverberation, we do not use the onset test. The block diagram of the single-source histogram algorithm is shown in Fig.~\ref{fig2:histogram} The theoretical steering vector for the first-order Ambisonics format is approximately true for up to $9$ kHz. Therefore we only search for single-source TF bins between bin $2$ and bin $384$, which correspond to $50$ Hz and $9000$ Hz respectively. 

\begin{figure}[tb]
\centering
\includegraphics[width=0.5\textwidth]{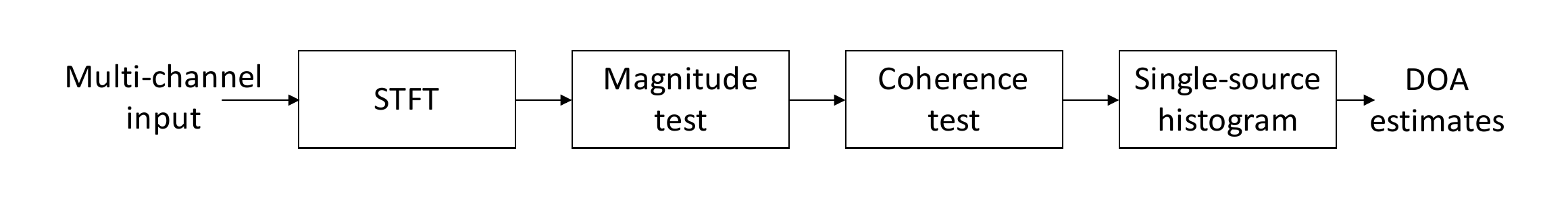}
\caption{Block diagram of single-source histogram algorithm}
\label{fig2:histogram}
\end{figure} 

We extend the $2$D DOA estimation in ~\cite{tho2014robust} to $3$D DOA estimation for DCASE 2019 task 3. The ranges of azimuth and elevation angles of the room impulse responses in task 3 are [$-180^{\circ}$, $180^{\circ}$), and [-$40^{\circ}$, $40^{\circ}$] respectively. The azimuth and elevation resolutions are $10^{\circ}$. After discretizing, the dimensions of the $2$D DOA histogram are $16$x$9$. From our observations, the estimated elevation angles have higher variation than the estimated azimuth angles. Therefore, we smooth the $2$D histograms using a $2$D Gaussian kernel with higher variance for elevation. Validation sets are used to find the best threshold to select the peaks on the DOA histograms. The evaluation metrics used for DOA task are angle error and frame recall that are defined in~\cite{adavanne2018doa_EUSIPCO}.  

\subsection{Combine SED and DOA estimations}
\label{sec:typestyle}

We use a set of rules to combine SED and DOA estimation at each frame. The SED results have higher precedence than the DOA results. Let denote $n_{sed}$ and $n_{doa}$ as the number of sound events detected by SED and DOA estimator respectively. Since the SELD dataset has maximum of $2$ overlapping events for each time frame, we limit $n_{sed}$ and $n_{doa}$ to $2$. Algorithm~\ref{algorithm} describes the combination rules. The limitation of this approach is that when SED and DOA estimator return $2$ sources, the DOAs and classes of the sound events have $50$\% chance of being mismatched. This mismatched issue will be studied in our future research. 

\begin{algorithm}
\caption{Combine SED and DOA estimation}
\label{algorithm}
\begin{algorithmic}[1]
\State For each time frame
\If {$n_{SED} == 0$}
	\State Return None
\ElsIf {$n_{SED} < n_{DOA}$}
	\State Assign all the DOAs to the sound event
\ElsIf {$n_{SED} == n_{DOA}$}
    \State Randomly assign one DOA for each sound event
\Else
	\State Look for additional DOAs in the neighbourhood frames
	\If {find addition DOAs so that $n_{SED} == n_{DOA}$ }
	    \State Randomly assign one DOA for each sound event
	\Else
	    \State Randomly ignore the extra sound event
	\EndIf
\EndIf
\State Return pairs of (sound event, DOA)
\end{algorithmic}
\end{algorithm}

\section{Experimental results and discussion}
\label{sec:majhead}

\subsection{SED results}

Table~\ref{table2: SED results} shows the SED results of the proposed CRNN-Resnet model and four variations using four-fold cross validation. We see that the model uses LSTM (variation $2$) is slightly better than CRNN-Resnet that uses GRU in term of error rate but not F$1$ score. In addition, adding spatial features as in variation $3$ and $4$ also improves error rate. Variation $4$ has an additional output and its performance is slightly lower than those of variation $3$. The model ensemble achieves the best error rate and F$1$ score as expected. The ensemble reduces the error rate by $0.06$ and increase the F$1$ score by $2.2$\% compared to the best single model. The number of parameter of the CRNN-Resnet and the ensemble are $1$ million and $5.6$ million respectively. 

\begin{table}\small
\caption {SED test results using development dataset}  \vspace*{5pt}
\centering
\label{table2: SED results}
\scalebox{0.8}{
\begin{tabular}{|c|c|c|}
\hline 
Model & ER & F1(\%) \\ 
\hline 
CRNN-Resnet & 0.250 & 87.3 \\ 
\hline 
CRNN-Resnet variation 1 & 0.264 & 85.0 \\ 
\hline 
CRNN-Resnet variation 2 & 0.252 & 85.4 \\ 
\hline 
CRNN-Resnet variation 3 & 0.230 & 86.2 \\ 
\hline 
CRNN-Resnet variation 4 & 0.233 & 86.0 \\ 
\hline 
Ensemble & \textbf{0.170} & \textbf{89.5} \\ 
\hline 
\end{tabular} 
}
\end{table}

\subsection{DOA results}

Fig.~3  shows the DOA spectra of the smoothed single-source histogram and MUSIC algorithm for a two-source case. The active source classes are \emph{drawer} and \emph{laughter} at ($-120^{\circ}$, $-10^{\circ}$) and ($100^{\circ}$, $-20^{\circ}$) respectively. The red cross and magenta circle markers show the ground truth and the estimated DOAs respectively. We see that the DOA estimates using the single-source histogram coincide with the ground truths while the DOA estimates using MUSIC algorithm do not. Table~\ref{table3: DOA_results} shows the four-fold cross validation results of the MUSIC and single-source histogram algorithms. The single-source histogram reduces the DOA error by $2^{\circ}$ and increases the frame recall by $4.9$\% compared to those of MUSIC. 

\begin{figure}[!tbp]
  \begin{subfigure}[b]{0.4\textwidth}
    \includegraphics[width=\textwidth]{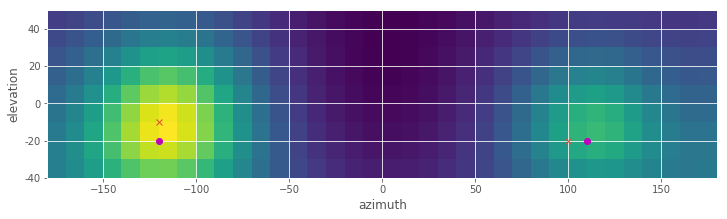}
    \caption{MUSIC spectrum}
    \label{fig:f1}
  \end{subfigure}
  \vfill
  \begin{subfigure}[b]{0.4\textwidth}
    \includegraphics[width=\textwidth]{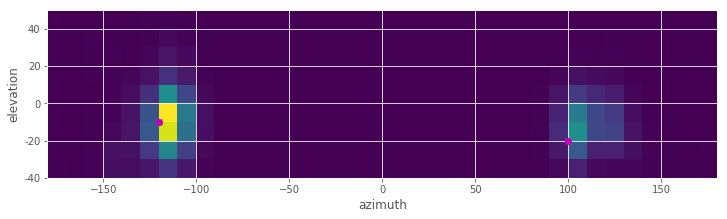}
    \caption{Smoothed single-source histogram}
    \label{fig:f2}
  \end{subfigure}
  \caption{Spatial spectrum of (a) MUSIC and (b) single-source histogram when there are $2$ sources}
  \label{fig3:doa_histogram}
\end{figure}

\begin{table}\small
\centering
\caption {DOA estimation results of MUSIC and single-source histogram algorithms}  \vspace*{5pt}
\label{table3: DOA_results}
\scalebox{0.8}{
\begin{tabular}{|c|c|c|}
\hline 
Algorithm & DOA Er & DOA FR (\%) \\ 
\hline 
MUSIC & 7.75 & 84.0 \\ 
\hline 
Single-source histogram & \textbf{5.15} & \textbf{88.9} \\ 
\hline 
\end{tabular} 
}
\end{table}

\subsection{SELD results}

We combine the SED prediction results from the single CRNN-Resnet and the ensemble model with the DOA estimations from the single-source histogram algorithm. We call these two system as single model and ensemble respectively. Table~\ref{table3: seld_results} shows the four-fold cross-validation results on the development dataset of the baseline system, the two-stage model proposed by Cao \emph{et al.}~\cite{cao2019polyphonic}, and our two proposed systems. The experimental results show that our proposed methods outperformed the baseline system by a large margin for SED error rate, F1 score, and DOA error. The single model achieves the best frame recall for DOA estimations. The ensemble achieves the best performance for SED error rate. The two-stage system achieves the best F$1$ score for SED. The DOA errors are similar for both the proposed single and ensemble system. The DOA frame recall of the ensemble is slightly lower than the single system's. The baseline system learns a CRNN model that jointly estimate sound events and DOAs from both magnitude and phase spectrogram. The two-stage system trains SED branch first, then transfer the CNN weights to DOA branch, and fine-tunes the DOA branch. Because SED and DOA estimation require different types of information from the microphone inputs, we do the SED and DOA estimation separately to maximize the performance of both tasks. The downside of our approach is the mismatch between sound classes and their corresponding DOAs when there are more than one sound event. However, because the evaluation metrics do not penalize this mismatch, we could not quantify this mismatch in both the baseline and our proposed algorithms.

\begin{table}\small
\centering
\caption {SELD performance using development set}  \vspace*{5pt}
\label{table3: seld_results}
\scalebox{0.80}{
\begin{tabular}{|c|c|c|c|c|}
\hline 
System & SED ER & SED F1(\%) & DOA Er & DOA FR \\ 
\hline 
baseline & 0.34 & 79.9 & 28.5 & 85.4 \\ 
\hline 
two-stage & 0.18 & \textbf{89.8} & 9.84 & 85.7 \\ 
\hline 
single model & 0.21 & 86.9 & 5.15 & \textbf{88.9} \\ 
\hline 
ensemble & \textbf{0.17} & 89.3 & \textbf{5.12} & 87.5 \\ 
\hline 
\end{tabular} 
}
\end{table}

Table~\ref{table4: eval_results} shows the evaluation results of the baseline system, the winning system, and our proposed single and ensemble system. Our proposed ensemble system has comparable performance on SED F$1$ score and DOA error. However, our DOA frame recall is much lower. The reason is that the single-source histogram tends to further divide a labeled sound event into smaller segments due to short pauses within the sound events. 

\begin{table}\small
\centering
\caption {SELD performance using evaluation set}  \vspace*{5pt}
\label{table4: eval_results}
\scalebox{0.80}{
\begin{tabular}{|c|c|c|c|c|}
\hline 
System & SED ER & SED F1(\%) & DOA Er & DOA FR \\ 
\hline 
baseline & 0.28 & 85.4 & 24.6 & 85.7 \\ 
\hline 
winning system & \textbf{0.08} & \textbf{94.7} & \textbf{3.7} & \textbf{96.8} \\ 
\hline 
single model & 0.15 & 91.1 & 5.6 & 89.8 \\ 
\hline 
ensemble & 0.11 & 93.4 & 5.4 & 88.8 \\ 
\hline 
\end{tabular} 
}
\end{table}

Table~\ref{table5: sub1} shows the performance of the proposed single system across all folds, overlaps, and impulse responses. Across $4$ folds, the error rate and the F$1$ score for SED task vary the most, while the DOA error and the frame recall are relatively constant. The system performance degrades when there are overlapping events. The SED error increases from $0.16$ to $0.24$ and the DOA frame recall drops from $96.0$\% to $81.8$\%. Across different rooms, the performance of SED is stable, while the DOA error has more fluctuation. These results show that overlapping sound event and the different room acoustics are the main challenges for the SELD task.

\begin{table}\small
\centering
\caption {Development results across folds, overlaps, and impulse responses of the proposed single system}  \vspace*{5pt}
\label{table5: sub1}
\scalebox{0.80}{
\begin{tabular}{|c|c|c|c|c|c|}
\hline 
 & ID & SED ER & SED F1 & DOA Er & DOA FR \\ 
\hline 
\multirow{4}{*}{\centering Fold} & 1 & \textbf{0.17} & \textbf{89.7} & 5.17 & \textbf{89.6} \\ 
                                 & 2 & 0.26 & 84.2 & 5.15 & 88.1 \\ 
                                 & 3 & 0.18 & \textbf{89.7} & 5.22 & 89.2 \\ 
                                 & 4 & 0.25 & 0.84 & \textbf{5.05} & 88.7 \\ 
\hline 
\multirow{2}{*}{\centering Overlap} & 1 & \textbf{0.16} & \textbf{90.4} & \textbf{4.66} & \textbf{96.0} \\ 
                                    & 2 & 0.24 & 85.0 & 5.41 & 81.8 \\ 
\hline 
\multirow{5}{*}{\centering Impulse Response} & 1 & \textbf{0.21} & 87.2 & \textbf{3.74} & 89.3 \\ 
                                             & 2 & \textbf{0.21} & \textbf{87.3} & 3.99 & \textbf{89.9} \\  
                                             & 3 & 0.22 & 86.4 & 7.17 & 88.2 \\ 
                                             & 4 & \textbf{0.21} & 87.2 & 4.69 & 89.4 \\ 
                                             & 5 & 0.22 & 86.4 & 6.03 & 87.7 \\ 
\hline 
\multicolumn{2}{|c|}{Total} & 0.21 & 86.9 & 5.15 & 88.9 \\ 
\hline 
\end{tabular} 
}
\end{table}

\section{Conclusion}
\label{sec:conclusion}

SELD is an interesting task with many real-life applications. Our experimental results show that a joint model for SED and DOA might be suboptimal. The separate SED and DOA estimation models achieve better performance on the DCASE task 3 dataset compared to the joint model in the baseline system. The advantage of using a separated parametric algorithm for DOA estimation is that it does not require a new training dataset for different microphone array configurations or different sound classes. The SED model can easily adapted to work on single channel and can leverage on many available datasets for SED. The drawback of our proposed system is the mismatch of the sound classes and the DOAs in multi-source cases. 

\bibliographystyle{IEEEtran}
\bibliography{refs}
%
%
%
%
%
%
%
%
%

\end{sloppy}
\end{document}